# Theory of Regulatory Compliance for Requirements Engineering

Ivan J. Jureta*†     Alberto Siena ‡     John Mylopoulos§     Anna Perini ¶

Angelo Susi ∥

October 26, 2018

**Abstract**

Regulatory compliance is increasingly being addressed in the practice of requirements engineering as a main stream concern. This paper points out a gap in the theoretical foundations of regulatory compliance, and presents a theory that states (i) what it means for requirements to be compliant, (ii) the compliance problem, i.e., the problem that the engineer should resolve in order to verify whether requirements are compliant, and (iii) testable hypotheses (predictions) about how compliance of requirements is verified. The theory is instantiated by presenting a requirements engineering framework that implements its principles, and is exemplified on a real-world case study.

*Fonds de la Recherche Scientifique – FNRS, Brussels, Belgium
†University of Namur, 8, rempart de la vierge, 5000 Namur, Belgium; `ivan.jureta@fundp.ac.be`.
‡FBK-Irst, via Sommarive 18, Trento, Italy; `siena@fbk.eu`.
§University of Trento, via Sommarive 14, Trento, Italy; `jm@disi.unitn.it`.
¶FBK-Irst, via Sommarive 18, Trento, Italy; `perini@fbk.eu`.
∥FBK-Irst, via Sommarive 18, Trento, Italy; `susi@fbk.eu`.

# Contents

# 1 Introduction

A legal norm (aka law, regulation) is a body of knowledge that acts both as a source of requirements and as a filter for requirements. Just as any requirement can be considered desirable or undesirable, it is also either compliant or not with relevant law, whereby compliance of a requirement refers informally to its alignment to law. Determining whether some set of requirements is compliant with relevant laws is therefore becoming a central issue in Requirements Engineering (RE). This was to be expected, for non-compliant requirements result in – quite literally – an illegal information system, thereby risking sanctions for their owners and users, among others. An RE project certainly cannot purport to be successfully executed if it fails to establish regulatory compliance.

Verifying and ensuring regulatory compliance is a difficult problem fought continually in and out of courtrooms with large budgets, over long periods of time. It should not be expected of the legislators who author laws to reduce either these budgets or time spans. Rather, the challenge is on the designers of information systems, and thus on requirements engineers who must at the very outset of the information systems engineering effort verify through a systematic process that requirements indeed comply with relevant laws. To do so, they need methods, and methods in turn require that we are clear on what regulatory compliance means in the first place.

This paper offers a theory of regulatory compliance and discusses its application via a real-world compliance problem. The theory starts from simple observations, prima facie evidence on how compliance is being engineered into systems, providing as a result the so-called compliance relation $\vdash_C$ between a requirements specification and set of norms. The compliance relation comes together with: (i) formal pragmatic conditions that requirements should satisfy in order to be considered compliant; (ii) a formal definition of the compliance problem, i.e., the problem that the engineer should resolve in order to verify that requirements are compliant; (iii) testable hypotheses about how regulatory compliance of requirements is verified.

The proposed framework is founded on argument-theoretic, as opposed to truth-theoretic, semantics. This means that compliance is not reduced to establishing that if requirements along with assumptions are "true" so is the norm with respect to which we are establishing compliance. Rather, compliance means that an argument that a given set of requirements complies with a given norm, is "acceptable" to stakeholders. The argumentation concepts used are adopted from [11], though in somewhat simplified form.

The rest of the paper is structured as follows. We start with an informal discussion of what it means for requirements to be compliant with some relevant law (§2). The three parts of the compliance theory for requirements are then presented and discussed in detail (§3). A requirements engineering framework is presented, which provides an implementation of the theory (§4–5). The theory is applied via an example of a real-world compliance problem (§6). The paper closes with the discussion of related work (§7), a summary of conclusions, and pointers for future research (§8).

# 2 Understanding Regulatory Compliance

The aim in this section is to introduce informally what it means for requirements to be compliant with some relevant law. Many details remain opaque here for the sake of simplicity; we formalize them later on.

It is intuitively clear that regulatory compliance of an information system concerns the relationship between the behavior of that running system and some relevant legal norms: the norms specify some constraints that must be satisfied and others that must not be violated, so that we can only say that the system complies if and only if no behavior of the system fails with regards to any of these two sets of constraints. Since behaviors should follow a specification thereof, compliance of requirements concerns the relationships between the specified behaviors of the system-to-be and a set of relevant legal norms.

Stated in the current common terminology of RE, *compliance concerns the relationships between the specification of the system-to-be and the set of relevant legal norms.* In turn, *the specification defines tasks that must be fullfilled, relates these tasks to agents who have been delegated these tasks, all in order to satisfy a set of goals.* Stated otherwise, we cannot discuss the compliance of the system-to-be without being concerned with the (i) tasks, (ii) delegations of tasks to agents, and (iii) the relationships between the pre/post conditions of the tasks and the legal norms.

Suppose that a doctor in a hospital subject to the USA Health Insurance Portability and Accountability Act applies has the goal to share patient information with colleagues. HIPAA states the following:

> SAFEGUARDS.—Each person described in section 1172(a) who maintains or transmits health information shall maintain reasonable and appropriate administrative, technical, and physical safeguards — (A) to ensure the integrity and confidentiality of the information; [...]" (42 USC 1320d-2, Sect.1173)

Given that there are potentially many different ways for a doctor in a hospital to share patient information with colleagues, we cannot verify regulatory compliance of a goal in relation the the safeguards mandated by HIPAA. We must instead look into all *solutions* of a goal, and in turn all solutions for all goals of the information system: compliance is verified between a *specification* and the legal norms.

What is a solution of a goal? One solution for the hospital doctor's goal is to hand over the patient's file to a nurse, and tell her to carry it to another doctor. Another solution may be for the doctor to go herself to another doctor. If patient files are stored on a hospital network, the doctor may simply share these files with her colleagues by changing the access rights for the patient file. The point is that there may be potentially many solutions to a goal, and all of these solutions need to be checked for compliance with legal norms. Any solution involves some agents, which have been delegated tasks so that if they execute tasks the goal will be satisfied.

To describe a solution is to record the tasks (i.e., what is done) and delegations (i.e., who does it) involved in the solution. Let $R$ be a model of requirements for a system-to-be, and let $Spec(R)$ be the specification for that system-to-be, i.e., the set of solutions for all goals in $R$. The next ingredient needed for the verification of compliance are the legal norms that are deemed relevant to the system in question.

It is important to distinguish between legal texts and a model of legal norms, for it is the latter that is used in the verification of compliance. Why is this so? While law does apply to specific domains, such as healthcare or financial industry, legal texts that codify law are rarely specific enough or written in a precise enough manner for them to leave no ambiguities, no vagueness, and in general, no openness in interpretation. Hence the need, as is commonly said, to "interpret law". A model of norms arises precisely out of an interpretation of a set of legal norms.

Suppose that we interpret some laws relevant to the system-to-be, and that this gives us a model of norms, $L$. Given $Spec(R)$ and $L$, we are still missing a third and final element, which we call *compliance assumptions*, $C$. The purpose of $C$ is to indicate how fragments of the specification relate to those of the model of norms. Returning to the exmaple, $C$ may indicate that if the doctor gives the patient file to a nurse and tells her to carry it to another doctor, then this guarantees that the integrity and confidentiality of patient's information is ensured. In other words, $C$ may say that some specific solution, if applied, will satisfy or not violate some (set of) legal norms.

Given $Spec(R)$, $L$, and $C$, we can sketch the condition that must be satisfied in order to say that $Spec(R)$ complies with $L$. Since we are interested in checking if $Spec(R)$ and $C$ satisfy $L$, then it seems appropriate to say that $Spec(R)$ complies with $L$ given $C$ if and only if the $L$ is the consequence of $Spec(R)$ and $C$: i.e., $Spec(R), C \mathrel{\mid\!\sim}_C L$. We assume that $Spec(R)$, $C$, and $L$ are sets of well-formed formulas (wffs) in some formalism. The relation $\mathrel{\mid\!\sim}_C$ is a logical consequence relation which we call *the compliance relation* and we formalize in the next section. Its salient traits are that it is nonmonotonic and that its proof theory is based on the concept of argument and the justification process.

Why are these two properties important — why not $Spec(R), C \vdash L$, where $\vdash$ is a standard monotonic consequence relation? That $\vdash$ is monotonic means (by the definition of the monotonicity property of $\vdash$) that if $Spec(R), C \vdash L$ then also $Spec(R'), C \vdash L$, where $Spec(R) \subseteq Spec(R')$. Informally, if $Spec(R), C \vdash L$ then whatever we add to $Spec(R)$ to obtain $Spec(R')$, the initial specification $Spec(R)$ will still be compliant with $L$. When a specification or laws change, it is necessary to verify compliance again. It is clear that we would very much prefer to work in settings in which $\mathrel{\mid\!\sim}_C$ were a monotonic consequence relation, for it would reduce the work necessary to verify compliance when new requirements are added and new laws are adopted, but this is simply not the case. Why arguments? We said above that $L$ is not a legal text, but a model of norms which is produced by the interpretation of a legal text. It follows that we can never really confidently say to be certain that a $Spec(R)$ is perfectly compliant with some law. We can instead defend the claim "$Spec(R)$ complies with $L$" against any counterarguments that may be offered. When new counterarguments become available, we are to consider whether we can defend $R$ against these new counterarguments, and if not, proceed to modify requirements so as to reestablish compliance. Verification of compliance thus requires a dialectical method, in which arguments that support the claim that there is compliance are defended against arguments that attack them. Stated otherwise, verification of compliance amounts to the justification of the claim "$Spec(R)$ complies with $L$" given some arguments that support and others that oppose that claim. Returning to the example, perhaps there is a stakeholder who may disagree that having a nurse carry a patient file to another doctor ensures the integrity and confidentiality of patient information: this stakeholder may suggest that if nurses are not aware of HIPAA regulations, they may take less care and so integrity or confiendtiality could be violated. The engineer will reestablish compliance of this solution by offering evidence against the stakeholder's argument: e.g., the engineer could indicate that nurses will receive training on HIPAA. The counterarguments that the engineer offers may lead to changes to any of $Spec(R)$, $C$, or $L$.

## 3 Theory of Regulatory Compliance

The intuitions outlined in the previous section are formalized here. We first give definitions for $R$, $Spec(R)$, $C$, and $L$ (§3.1). We then formalize *compliance frameworks* used to verify compliance, by giving their formal semantics and proof theory, thereby defining $\vdash_C$ (§3.2). We use compliance srtructures to define the compliance problem (§3.3) and testable hypotheses about how regulatory compliance is verified in relation to requirements (§3.4).

### 3.1 $R$, $Spec(R)$, $C$ and $L$

**Requirements Model $R$.** Requirements that are elicited, discussed, refined, or otherwise manipulated over the course of RE are recorded in a requirements model $R$. *$R$ amounts to a set of well-formed sentences of a modeling language $\mathcal{R}$*. Most current modeling languages in RE allow us to state propositions and combine them into sentences well-formed according to the grammar of the language. What is characteristic of RE languages is that the propositions will be sorted: e.g., the proposition *Patient data is available before diagnosis* may be an instance of the *goal* concept, or a postcondition of an instance of the *task* concept. Since $R$ describes a socio-technical setting, the instances of the various concepts will be associated to agents; e.g., the task of making patient data available before diagnosis may be a goal assigned to a doctor, who may then delegate it to another agent, and so on. $\mathcal{R}$ will consequently have at least the sets $P_R$, $A_R$, $S_R$ of symbols referring to, respectively, propositions, agents, and sorts, along with a function $\mathcal{P} : P_R \longrightarrow S_R \times A_R$ to relate every proposition to a sort and an agent. The modeling language will define a grammar over $(S_R, P_R, A_R)$ and at least a consequence relation so that we can compute logical consequences of well-formed sentences in the language. It may well have also a semantic entailment relation, but its presence or absence does not change our discussion below.

**Specification $Spec(R)$.** $R$ contain goal-propositions, i.e., $(g, p, a)$ where $g \in S_R$, $p \in P_R$ and $a \in A_R$, with $\mathcal{P}_R(p) = (g, a)$. Refinements and delegations of this or any other goal can be fairly elaborate: e.g., suppose that the goal in question is and-refined onto three goals, and that each of the subgoals can be delegated to either of three agents, so that if we assume one agent can satisfy all three of these goals, then there we have nine different ways of satisfying $(g, p, a)$. Moreover, every one of these agents can know of different sets of tasks that can satisfy the goal she is delegated, and she may further delegate the execution of these tasks to other agents. This point is made in Figure 1: it shows that $a$'s goal $p$ is and-refined onto three goals, that these three goals can be delegated in nine ways, and that each of the agents who is delegated one or more of these goals also has options regarding what tasks to delegate to whom. This merely reflects the reality of human organizations, in which there are many ways of realizing the same aim, depending on who is being delegated what tasks.

We say informally that a *solution* of a goal is the joint behavior of people and machines within a socio-technical system, which ends up bringing about a state of affairs in which the goal holds. A solution satisfies a goal. Within $R$, there can be many different solutions of the goal $(g, a, p)$, any one of which involves agents and their execution of tasks, i.e., tasks delegated to agents. Figure 1 shows two solutions, one having the task delegations $(t, q_1, q_2, a_4)$, $(t, r_3, r_4, a_7)$ and $(t, s_1, s_2, a_8)$ (white boxes in Figure 1), and the other having $(t, q_3, q_4, a_5)$, $(t, r_1, r_2, a_6)$ and $(t, s_3, s_4, a_9)$ (black boxes in Figure 1). Note that a every task has two propositions, the left stating preconditions and the right giving postconditions. The specification $Spec(R)$ will be a part of the requirements model $R$, and will contain a set of solutions for a set of parent goals (i.e., goals that are not refinements of other goals) from $R$.

**Norm Model $L$.** A norm model $L$ is a set of well-formed sentences written in a modeling language $\mathcal{L}$. Similarly to $\mathcal{R}$, there will be a modeling language in which the legal expert and/or the engineer record the interpretation of legal text. $\mathcal{L}$ will have a grammar that governs the formation of well-formed sentences over sets $P_L$, $R_L$, $S_L$ of symbols for, respectively propositions, roles, and sorts. We have roles in $\mathcal{L}$ instead of agents, as law typically speaks not of particular agents, but of roles (e.g., patient, physician, buyer, seller, and so on). Every sort symbol will refer to a concept in the legal ontology used in $\mathcal{L}$. E.g., if there is the concept of *right* in the chosen legal ontology, then there is bound to be a sort, which is associated to all propositions that capture what is understood as a right in the legal domain. Since rights are attributed to legal subjects (e.g., individuals) via roles, a model of legal norms needs the function $\mathcal{P}_L : P_L \longrightarrow S_L \times R_L$ to map every proposition from law to a sort and a role. Triples $(s \in S_L, p \in P_L, r \in R_L)$, such that $\mathcal{P}_L(p) = (s, r)$ are called *normative propositions* (NPs). In practice, an NP will be obtained from a sentence or a paragraph in a regulatory text; consider again for example the passage from HIPAA on SAFEGUARDS, quoted earlier (§2). That passage can be interpreted to give an NP, in which there is one role, *Person maintaining or transmitting health information*, a proposition *Integrity and confidentiality of the health information is ensured*, and a sort to indicate the what legal concept (e.g., *obligation*) this proposition instantiates.

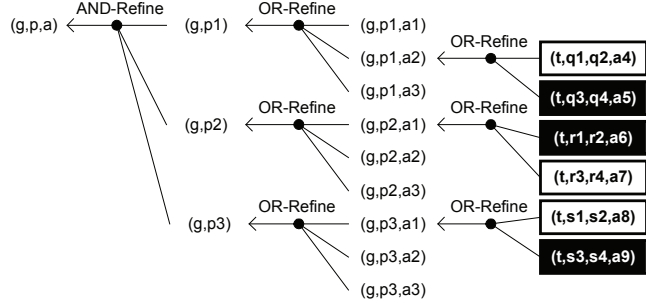

**Figure 1:** Two solutions of the goal $(g, p, a)$.

**Compliance Assumptions** $C$. Given $Spec(R)$ and $L$, how are they related? The relations between individual (fragments of) $Spec(R)$ and $L$ cannot be obtained immediately from the two of them: the very purpose of a method for the verification of compliance is to formulate a set $C$ of *compliance assumptions*. What a compliance assumption $c \in C$ does is that it relates propositions or well-formed sentences from $Spec(R)$ to those from $L$. How does this happen in practice? Suppose that $\phi_R$ and $\psi_L$ are two well-formed sentences from, respectively $Spec(R)$ and $L$, so that, e.g., $c$ may be $c \equiv \phi_R \to \psi_L$, where $\to$ is the implication connective. If the framework for reasoning about compliance has modus ponens as an inference rule, then we apply modus ponens to $(\phi_R, c)$ and conclude that $\psi_L$. Suppose that we have a $\phi_R = q \wedge q'$, where "$q$: *The patient's file is received for transfer from a doctor*" and "$q'$: *The patient file has been transferred to a doctor*" and both $q$ and $q'$ have *task* for one sort and *hospital nurse* for the other. Stated otherwise, a nurse is delegated the task to carry the patient's file between doctors. Let "$\psi_L$: *Integrity and confidentiality of the health information is ensured*". In this case, a compliance assumption $c \equiv \phi_R \to \psi_L$ relates these sorted well-formed sentences, claiming that if the hospital nurse ensures that $\phi_R$, then $\psi_L$ holds, i.e., we are compliant with $\psi_L$.

## 3.2 Arguing Compliance

To encode this conclusion that compliance is verified through an process of argumentation, we define $\vdash_C$ as a consequence relation within an argumentation framework. In doing so, a claim that "$R$ complies with $L$" is at best a tentative assertion, open to revision as requirements, legislation, and interpretation of legislation change, and as new counterarguments are voiced which a given compliance framework cannot defend itself against. Hence the need for a *continual* verification of compliance, clearly particularly relevant given that change is more the rule than the exception in information systems engineering.

Suppose that we have $Spec(R)$, $L$, and $C$, and that they are all well-formed formulas of a sorted propositional logic. Stated otherwise, we say that while $\mathcal{L}$ and $\mathcal{R}$ may be different modeling languages, they differ only in sorts and in that the former maps propositions to sorts and *roles* and the latter to sorts and *agents*. In practice, $\mathcal{R}$ and $\mathcal{L}$ may both be sorted propositional logics with the same proof theory, and thus the same consequence relation $\vdash$.[1] We let $\Gamma$ be an arbitrary (potentially inconsistent) set of well-formed formulas in

---

[1] If $\mathcal{R}$ and $\mathcal{L}$ are two different logics, e.g., the former a linear-temporal first-order logic, and the latter a deontic logic, then it is necessary to find a joint framework in which the well-formed formulas of these two share a proof theory. This is a problem of

the logic in which $Spec(R)$, $L$, and $C$ are written.

**Definition 3.1.** *(Compliance) argument.* $\Gamma' \subseteq \Gamma$ *is called a compliance argument for* $\phi \in \Gamma$, *written* $\langle \Gamma', \phi \rangle$ *iff:*

1. *$\phi$ is a consequence of $\Gamma'$: $\Gamma' \vdash \phi$;*

2. *$\neg\phi$ is not a consequence of $\Gamma'$: $\Gamma' \not\vdash \neg\phi$;*

3. *$\Gamma'$ has only the necessary and sufficient well-formed sentences for the derivation of $\phi$: $\forall \Gamma'' \subset \Gamma'$, $\Gamma'' \not\vdash \phi$.*

*Example* 3.2. Let $\Gamma' = \{\phi_R, \phi_R \to \psi_L\}$, then $\Gamma'$ is an argument for $\psi_L$.

**Definition 3.3.** *Compliance relation $\mathrel{\mid\!\sim}_C$. Let $A(\Gamma)$ be the set of all arguments from $\Gamma$. $\phi$ is compliant with $\Gamma$, written $\Gamma \mathrel{\mid\!\sim}_C \phi$ iff:*

1. *there is an argument for $\phi$ in $A(\Gamma)$, and*

2. *there is no argument for $\neg\phi$ in $A(\Gamma)$.*

*Example* 3.4. Let $\Gamma = \{\phi_R, \phi_R \to \psi_L, \phi_R \to \gamma_L\}$, then $\Gamma \mathrel{\mid\!\sim}_C \psi_L$ and $\Gamma \mathrel{\mid\!\sim}_C \gamma_L$.

**Proposition 3.5.** $\mathrel{\mid\!\sim}_C$ *is nonmonotonic.*

*Proof.* (By contradiction.) If $\Gamma = \{\phi\}$, then $\Gamma \mathrel{\mid\!\sim}_C \phi$, so that if $\neg\phi$ is added to $\Gamma$, $\Gamma \cup \{\neg\phi\} \mathrel{\mid\!\not\sim}_C \neg\phi$. □

The following proposition says that nonmonotonicity appears only when $\Gamma$ is inconsistent w.r.t. $\phi$, and this is the case when both $\Gamma \vdash \phi$ and $\Gamma \vdash \neg\phi$.

**Proposition 3.6.** *If $\Gamma$ is consistent w.r.t. $\phi$, then $\Gamma \vdash \phi$ iff $\Gamma \mathrel{\mid\!\sim}_C \phi$.*

*Proof.* Obvious: if $\Gamma$ is consistent w.r.t. $\phi$, then $\Gamma \vdash \phi$ and $\Gamma \not\vdash \neg\phi$. If $\Gamma \mathrel{\mid\!\sim}_C \phi$ then by the definition of $\mathrel{\mid\!\sim}_C$, $\Gamma \mathrel{\mid\!\not\sim}_C \neg\phi$, so that there is an argument in $\Gamma$ for $\phi$ and there is no argument in $\Gamma$ for $\neg\phi$. As there is no argument in $\Gamma$ for $\neg\phi$, there is no derivation from $\Gamma$ to $\neg\phi$. □

### 3.2.1 Compliance Frameworks

**Definition 3.7.** *Compliance framework. A compliance framework is a tuple $\mathcal{C} = (Spec(R), L, C, \mathrel{\mid\!\sim}_C)$, where $R$ is a requirements model, $L$ a norm model, $C$ a set of compliance assumptions, and $\mathrel{\mid\!\sim}_C$ the compliance relation.*

It is not difficult to see that $\mathrel{\mid\!\sim}_C$ taken alone is defined in the same way as Benferhat, Dubois, and Prade's "argumentative consequence" [4], which itself follows a long line of work on argumentation in artificial intelligence, and in particular Simari and Loui's contributions [14], and Dung's seminal synthesis [7]. This alignment with available results in AI is particularly useful, since it is now possible in a straighforward manner to draw on fundamental contributions on the analysis of argumentation frameworks in order to define the semantics and proof theory for compliance frameworks. In order to state precisely the relation between a compliance framework and a (Dung's) argument framework, we need the standard argument attack relation.

**Definition 3.8.** *Argument attack relation ($\xrightarrow{A}$). Let $A(\Gamma)$ be the set of all arguments constructed from $\Gamma$, and let $\langle \Gamma_1, \phi \rangle$ and $\langle \Gamma_2, \psi \rangle$ be two arguments in $A(\Gamma)$, then $\langle \Gamma_1, \phi \rangle$ attacks $\langle \Gamma_2, \psi \rangle$, written $\langle \Gamma_1, \phi \rangle \xrightarrow{A} \langle \Gamma_2, \psi \rangle$ iff the former either rebutes or undercuts the latter argument, where:*

- $\langle \Gamma_1, \phi \rangle$ *rebutes* $\langle \Gamma_2, \psi \rangle$ *iff* $\psi \equiv \neg\phi$;

- $\langle \Gamma_1, \phi \rangle$ *undercuts* $\langle \Gamma_2, \psi \rangle$ *iff* $\exists \epsilon \in \Gamma_2$ *s.t.* $\phi \equiv \neg\epsilon$.

---

logic interoperability. Interconnections between formalisms can be characterized using category theory (e.g., [8]), whereby the relationships are established using functors; another way is to give common semantics to the different formalisms and translate them all into a common style of predicate logic, so that the relationships between them can be described as translations through the common syntax and semantics [17]. We will remain in the cost-effective case in this paper, which does not require that formalism be interconnected.

*Example* 3.9. If $\langle\{\phi, \phi \to \neg\psi\}, \neg\psi\rangle$ and $\langle\{\gamma, \gamma \to \psi\}, \psi\rangle$, then the former rebutes the latter, and the latter rebutes the former argument.

*Example* 3.10. If $\langle\{\phi, \phi \to \neg\psi\}, \neg\psi\rangle$ and $\langle\{\gamma, \gamma \to \neg\phi\}, \neg\phi\rangle$, then the latter undercuts the former argument.

The relation between a compliance framework and a (Dung's) argumentation framework is given by the following theorem.

**Theorem 3.11.** *There is for every compliance framework $\mathcal{C} = (Spec(R), L, C, \vdash_C)$ a Dung's argumentation framework $(A(\Gamma), \xrightarrow{A})$ such that:*

1. *$\Gamma = Spec(R) \cup L \cup C$ and $A(\Gamma)$ is the set of all arguments from $\Gamma$, and*

2. *$\xrightarrow{A}$ is the argument attack relation.*

*Proof.* (Trivial, from definitions.) An argumentation framework is by definition [7] a pair $(A(\Gamma), \xrightarrow{A})$, where $A(\Gamma)$ is a set of arguments and $\xrightarrow{A}$ is a binary relation on $A(\Gamma)$, i.e., $\xrightarrow{A} \subseteq A(\Gamma) \times A(\Gamma)$. Dung leaves $\xrightarrow{A}$ unspecified and an argument an abstract entity without considering its internal structure. Our compliance framework populates an argumentation framework, i.e., gives $A(\Gamma)$ and $\xrightarrow{A}$ is defined in terms of the components of the compliance framework. In doing so, every compliance framework populates an (Dung's) argumentation framework. □

### 3.2.2 Semantics of Compliance Frameworks

Theorem 3.11 allows us to give to a compliance framework the semantics of the argumentation framework that the compliance framework defines, and later to a proof theory that is particularly relevant from the standpoint of defining a method for the verification of compliance.

Let $a(\phi), b(\phi), c(\phi)$, primed or indexed be arguments for their respective parameters, e.g., $a(\phi) = \langle \Gamma', \phi \rangle$. The semantics follow Dung's notion of extensions of an argumentation framework, letting us state precisely what it means to say that an argument for, e.g., a legal norm in a compliance framework is acceptable. An extension is simply a set of arguments that satisfies some properties. Definitions 3.12–3.14 closely follow Dung's definitions [7] and Amgoud and Prade's rewriting [3] of Dung's definitions.

**Definition 3.12.** *Conflict-free set of arguments; Acceptable argument.*

- *A set $S$ of arguments is conflict-free iff $\nexists a(\phi), b(\psi) \in S$ s.t. $a(\phi) \xrightarrow{A} b(\psi)$.*

- *$a(\phi) \in A(\Gamma)$ is acceptable w.r.t. a set $S \subseteq A(\Gamma)$ of arguments iff $\forall b(\psi) \in A(\Gamma)$, if $b(\psi) \xrightarrow{A} a(\phi)$, then $\exists c(\epsilon) \in S$ s.t. $c(\epsilon) \xrightarrow{A} b(\psi)$.*

**Definition 3.13.** *Acceptability semantics. Let $(A(\Gamma), \xrightarrow{A})$ be an argumentation framework and $S \subseteq A(\Gamma)$ a conflict-free set of arguments.*

- *$S$ is an admissible extension iff every $a(\phi) \in S$ is acceptable w.r.t. $S$.*

- *$S$ is a preferred extension iff $S$ is a maximal (w.r.t. set $\subseteq$) admissible set.*

- *$S$ is a stable extension iff it is a preferred extension that $\xrightarrow{A}$-attacks every argument in $A(\Gamma) \setminus S$.*

Given the definition of extensions, a status can be assigned to an argument.

**Definition 3.14.** *Argument status. Let $(A(\Gamma), \xrightarrow{A})$ be an argumentation framework and $\mathcal{E}_1, \ldots, \mathcal{E}_x$ its extensions. Let $a(\phi) \in A(\Gamma)$:*

- *$a(\phi)$ is skeptically acccepted iff $\forall 1 \leq i \leq x$, $a(\phi) \in \mathcal{E}_i$;*

- *$a(\phi)$ is credulously accepted iff $\exists \mathcal{E}_i$ s.t. $a(\phi) \in \mathcal{E}_i$;*

- *$a(\phi)$ is rejected iff $\nexists \mathcal{E}_i$ s.t. $a(\phi) \in \mathcal{E}_i$.*

It is quite evident why the above are relevant for the verification of compliance. Given a compliance framework $\mathcal{C} = (Spec(R), L, C, \mathrel{\vdash}_C)$ and its associated argumentation framework $(A(\Gamma), \stackrel{A}{\longrightarrow})$, we may want to design $R$ so that $\forall \psi_L \in L$, there is an argument $a(\psi_L) \in A(Spec(R) \cup C)$ and $a(\psi_L)$ is skeptically accepted.[2] If every argument for every legal norm is skeptically accepted, then we have indeed defended all arguments given for all legal norms, i.e., the compliance framework ensures compliance.

We can now relate extensions and the compliance relation $\mathrel{\vdash}_C$ via Theorem 3.15, which is a useful result: it will allow us later to define the compliance problem as a problem of ensuring that a set of arguments associated to a compliance framework is skeptically accepted.

**Theorem 3.15.** *Let $\mathcal{C} = (Spec(R), L, C, \mathrel{\vdash}_C)$ be a compliance framework, $(A(\Gamma), \stackrel{A}{\longrightarrow})$ its associated argumentation framework, and $\mathcal{E}_1, \ldots, \mathcal{E}_x$ the extensions of the given argumentation framework. If $\Gamma \mathrel{\vdash}_C \phi$ then $\exists a(\phi) \in A(\Gamma)$ such that $a(\phi) \in \bigcap_{i=1}^{x} \mathcal{E}_i$, i.e., $a(\phi)$ is skeptically accepted.*

*Proof.* Because an argument can attack another argument in two ways (i.e., by rebutting or undercutting), the proof has two parts:

1. The first part is to show that if $\Gamma \mathrel{\vdash}_C \phi$, then there is no extension that has $a(\neg\phi)$, so that no extension contains an argument that rebuts $a(\phi)$. This is obvious: if $\Gamma \mathrel{\vdash}_C \phi$, then $\exists a(\phi) \in A(\Gamma)$ and $\nexists a(\neg\phi) \in A(\Gamma)$. Since $\forall 1 \leq i \leq x$, $\mathcal{E}_i \subseteq A(\Gamma)$, then clearly $\forall 1 \leq i \leq x$, $a(\neg\phi) \notin \mathcal{E}_i$, i.e., no extension contains $a(\neg\phi)$. Stated otherwise, if $\Gamma \mathrel{\vdash}_C \phi$, then $a(\neg\phi)$ is rejected.

2. The second part is to show that if $\Gamma \mathrel{\vdash}_C \phi$, then there is an argument $a(\phi) \in A(\Gamma)$ and every argument $a(\psi) \in A(\Gamma)$ that undercuts $a(\phi)$ is rejected. This second part of the proof is by contradiction. Suppose that there is one argument $a(\psi) \in A(\Gamma)$ such that $\psi \neq \neg\phi$ (i.e., $a(\psi)$ does not rebut $a(\phi)$) and $a(\psi) \stackrel{A}{\longrightarrow} a(\phi)$, so that $a(\psi)$ undercuts $a(\phi)$. Let $a(\phi) = \langle \Gamma', \phi \rangle$ and $a(\psi) = \langle \Gamma'', \psi \rangle$, and since $a(\phi), a(\psi) \in A(\Gamma)$, then $\Gamma' \subseteq \Gamma$ and $\Gamma'' \subseteq \Gamma$. Note also that by the definition of an undercut attack, if $a(\psi)$ undercuts $a(\phi)$, then $\neg\psi \in \Gamma'$ and thus $\neg\psi \in \Gamma$. There are three cases to consider:

    (a) $a(\psi)$ is skeptically accepted, i.e., $a(\psi) \in \bigcap_{i=1}^{x} \mathcal{E}_i$: Since $a(\phi) \in A(\Gamma)$ and $\neg\psi \in \Gamma'$, $\neg\psi$ is necessary for the derivation of $\phi$. If the argument for $\psi$, i.e., $a(\psi)$ is skeptically accepted, then it is in every extension possible to derive $\psi$ from $\Gamma$. If $\psi$ can be derived in every extension, then $\neg\phi$ can be derived in at least one extension, and thus $a(\neg\phi) \in \bigcup_{i=1}^{x} \mathcal{E}_i$. This contradicts the assumption of $\Gamma \mathrel{\vdash}_C \phi$, because by definition of $\mathrel{\vdash}_C$, $a(\neg\phi) \notin A(\Gamma)$, and thus $\forall 1 \leq i \leq x$, $a(\neg\phi) \notin \mathcal{E}_i$.

    (b) $a(\psi)$ is credulously accepted, i.e., $\exists \mathcal{E}_i$ such that $a(\psi) \in \mathcal{E}_i$: Suppose that there is an extension $\mathcal{E}_i \subseteq A(\Gamma)$ such that $a(\psi) \in \mathcal{E}_i$. Now, $\neg\psi$ is necessary for the derivation of $\phi$. It follows that there must be at least one other extension $\mathcal{E}_j \subseteq A(\Gamma)$, such that $a(\neg\phi) \in \mathcal{E}_j$. This contradicts the assumption of $\Gamma \mathrel{\vdash}_C \phi$, because by definition of $\mathrel{\vdash}_C$, $a(\neg\phi) \notin A(\Gamma)$, and thus $\forall 1 \leq i \leq x$, $a(\neg\phi) \notin \mathcal{E}_i$.

    (c) It follows from (a) and (b) above that $a(\psi)$ is rejected, i.e., $\nexists \mathcal{E}_i$ such that $a(\psi) \in \mathcal{E}_i$.

The conclusion of the proof is that if $\Gamma \mathrel{\vdash}_C \phi$, then $\exists a(\phi) \in A(\Gamma)$ and none of the attackers on $a(\phi)$ is in any of the extensions of $(A(\Gamma), \stackrel{A}{\longrightarrow})$. □

### 3.2.3 Proof Theory for Compliance Frameworks

The proof theory can now be defined for the semantics. The proof theory defines a practially applicable approach to the evaluation of the status of a given argument, and thereby of the membership of that argument in extensions. The proof theory is inspired by Simari and Loui's justification process [14], closely related to Amgoud and Cayrol's notion of a *dialogue* [2], itself arising out of the modeling of argumentation in the legal domain.

**Definition 3.16.** ***Justification process.*** *The justification process consists of recursively defining and labeling a dialectical tree $T(a(\phi))$ as follows:*

1. *A single node containing the argument $a(\phi)$ with no attackers is by itself a dialectical tree for $a(\phi)$. This node is also the root of the tree.*

---
[2]We say "may", because it can happen that a violation of legislation is preferred by the stakeholders of the information system, which can happen, e.g., when they consider their private benefits of that violation to be superior to the costs of ensuing sanctions.

2. *Suppose that $a(\psi_1), a(\psi_2), \ldots, a(\psi_n)$ each attacks $a(\phi)$. Then the dialectical tree $T(a(\phi))$ for $a(\phi)$ is built by placing $a(\phi)$ at the root of the tree and by making this node the parent node of roots of dialectical trees rooted respectively in $a(\psi_1), a(\psi_2), \ldots, a(\psi_n)$.*

3. *When no other argument in $A(\Gamma)$ can be added to the tree, label the leaves of the tree accepted (A). For any inner node, label it undefeated if and only if every child of that node is a rejected (R) node. An inner node will be a rejected node if and only if it has at least one A node as a child. Do the furth step below after the entire dialectical tree is labeled.*

4. *$a(\phi)$ is a justification iff the node $a(\phi)$ is labelled A.*

It is clear that if $a(\phi)$ is justified, then it is skeptically accepted.

## 3.3 Compliance Problem

**Definition 3.17. Compliance problem.** *Given $L$, find a compliance framework $\mathcal{C} = (Spec(R), L, C, \mathord{\vdash}_C)$ such that:*

$$Spec(R), C \mathrel{\vdash}_C L$$

Theorem 3.15 lets us reformulate the compliance problem as one the employs the justification process. Namely, given a norm model $L$, find a compliance framework $\mathcal{C} = (Spec(R), L, C, \mathord{\vdash}_C)$ such that: $\forall \psi_L \in L$, $\exists a(\psi_L) \in A(Spec(R) \cup C)$ and $a(\psi_L)$ is skeptically accepted, or in other words, the dialectical tree $T(a(\psi_L))$ indicates that $a(\psi_L)$ is a justification.

## 3.4 Testable Hypotheses

The rewriting of the compliance problem in terms of justification indicates that the verification of compliance will amount to the construction of the dialectical tree for every argument $a(\psi_L)$, $\psi_L \in L$. The theory of compliance thus hypothesizes that the verification of compliance will happen via the collection and confrontation of arguments, and that requirements will be considered as compliant when all arguments for norms, i.e., all $a(\psi_L)$ are evaluated as justifications via the justification process. The engineer and the stakeholders will offer $R$, $L$, and $C$, and perhaps arguments that support the given legal interpretation $L$ and the compliance assumptions $C$. In turn, some stakeholders may disagree with the content of any of $R$, $L$, or $C$, offering ccounterarguments, which will result in the lack of compliance. The engineer will in turn need to reestablish compliance by counterarguing and perhaps modifying $R$, $L$ or $C$. It follows that the extent to which compliance will be verified will depend strongly on the availability of arguments that attack every $a(\psi_L \in L)$. The availability of such attacking arguments in turn depends on the activity/passivity of the stakeholders who are attentive to the compliance of the requirements, which in turn means that in absence of such stakeholders, dialectical trees are bound to be shallow. In a summary, the testable hypotheses are as follows.

**Hypothesis 3.1.** *Verification of regulatory compliance of requirements proceeds as a justification process.*

**Hypothesis 3.2.** *Because of Hypothesis 3.1, the compliance or noncompliance of a requirements model $R$ depends on the availability of arguments against every $a(\psi_L \in L)$.*

**Hypothesis 3.3.** *Because of Hypothesis 3.2, the more easily accessible are the attacking arguments, the more resources (time, money, expertise) will be needed the verification of compliance.*

**Hypothesis 3.4.** *Because of Hypothesis 3.3, in absence of stakeholders who can voice the attacking arguments, the compliance of a requirements model will be verified over shallow dialectical trees.*

# 4 The *Nómos* modeling language

*Nómos* is a goal-oriented modeling framework, intended to support analysts in building models of law-compliant requirements as defined in section 3.3. *Nómos* is conceived as an extension of the *i\** modeling language [16], introducing a set of concepts to model laws. This way, it allows for a coherent representation of both $R$ and $L$.

Requirements in $R$ are represented in *Nómos* via $i^*$ concepts such as goals, which are states of affairs desired by stakeholders, represented as actors. Goals can be And- or Or-decomposed into sub-goals, and can be operationalized by means of tasks, which represent the solution for goals. Additionally, actor can delegate to each other the achievement of goals and the execution of tasks. The $i^*$ language fits the definition given in section 3.1 of a requirements modeling language: the statements describing the intentionals concepts (goals, tasks, resources) are in the set $P_R$; the sorts are $S_R = \{goal, softgoal, task, resource\}$, whereas $i^*$ actors form the set $A_R$.

Similarly, elements of $L$ are represented by giving an explicit supports the aforementioned concept of normative proposition (NP). NPs are propositions with a normative semantics, and contain information concerning: (i) the subject, who is addressed by the NP itself; (ii) the legal modality (i.e., whether it is a duty, a privilege and so on); and (iii) the description of the object of such modality (i.e., what is actually a duty or privilege). The legal modality is one of the 8 elementary rights, classified by Hohfeld [10] as: *Privilege*, which is the entitlement for a person to discretionally perform an action, regardless of the will of others who may not claim him to perform that action, and have therefore a *No-claim*; *Claim*, which is the entitlement for a person to have something done from another person, who has therefore a *Duty* of doing it; *Power*, which is the (legal) capability to produce changes in the legal system towards another subject, who has the corresponding *Liability*; and *Immunity*, which is the right of being kept untouched from other performing an action, who has therefore a *Disability*. So the sorts in $L$ are the elements of $S_L = \{duty, claim, privilege, noclaim, power, liability, immunity, disability\}$; the subjects of the NPs form the set $R_L$; and the object of the NPs form the set $P_L$

Elements of $R$ can be associated to elements of $L$ by means of 'realization' relations, which form the set $C$. When a goal realizes a NP, it means that the goal is either known to be, or intended to be, conform with the NP. For example, the *Nómos* model of Figure 2 depicts the actor Patient that depends on Hospital to access the healthcare services. Hospital provides this goal for the patient by fulfilling the goal (among possible others) of booking internally the service (goal Book service; according to the notation introduced in 2, $p_1$ = "Book healthcare service", $a_1$ = "Hospital", $s_1$ = 'goal'). The Hospital, in turn, depends on its internal Booking Service (BS) to book the service through the EPR. The BS asks the patients his/her data, accesses the EPR system, and inserts the data. The Figure shows that the Hospital has towards the Patient the duty to keep his/her Personal Health Information (PHI) closed. It is also shown – by means of the arrow between the goal "Book service" and the NP "Don't disclose PHI" – that by fulfilling that goal the Hospital is compliant with the NP. The next section illustrates *how* we built this model, while the subsequent section explains *why* the Hospital is considered to be compliant.

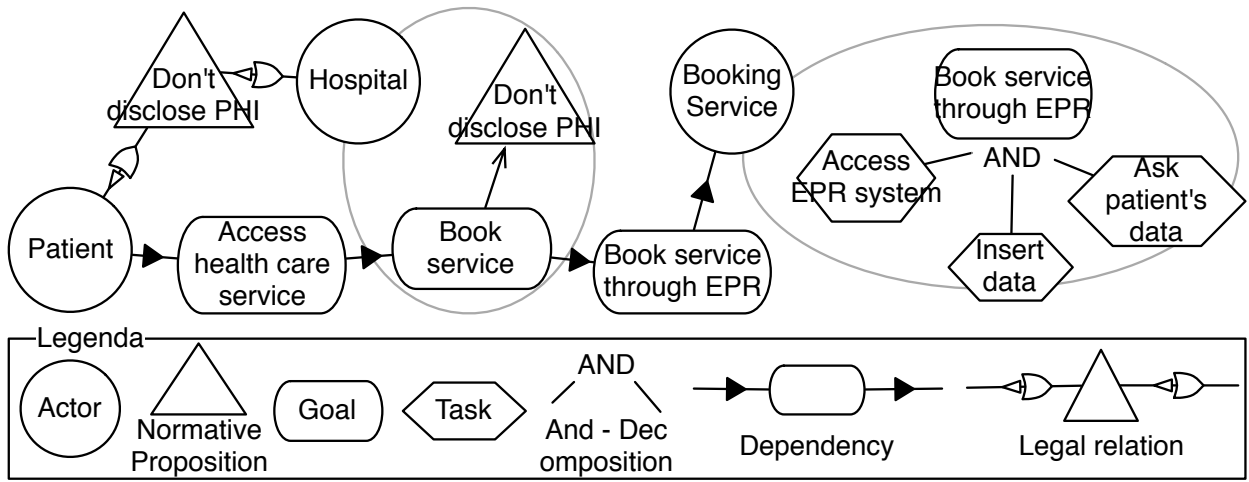

**Figure 2:** An example of a *Nómos* model.

# 5  The compliance process

The *Nómos* language allows for building models of requirements that comply with law according to the theory described in section 3. This is done through a process, which takes into consideration: (i) the production of a compliance solution; (ii) the argumentation supporting (or attacking) such solution; and (iii) the algorithm for making conclusions from the argumentation. The three corresponding phases form an iterative process, in which argumentation moves from the modeled solution and may require to change it, until the algorithm allows for deducing compliance acceptance.

**Phase 1. Compliance modeling**  It basically consists in the application of the *Nómos* modeling process. The methodological aspects of the process are described in detail in [13]: the process is basically comprised by various steps, each of them starting by arguing a compliance issue, and ending with agreement on the solution for that issue. For the purpose of this paper, we will focus on those parts of the process, which deal with the actual establishment of the compliance condition, without considering law modeling aspects mentioned in [13].

1. *Identify applicable laws.* Each NP addresses a subject. If a certain actor of the domain is subject to the NP, the NP is applicable to it and it has to comply w.r.t. the NP. For example, if the law addresses a "Covered Entity" (CE), and a stakeholder of the domain is a Hospital, we have to decide if the Hospital is (or not) a CE. In the first case, the NPs of the law apply to the Hospital, and it has to be compliant. If an actor is agreed to be a certain legal subject, this information is added to the argumentation tree.

2. *Identify affected goals.* Each NP applicable to a given actor can affect the actor's strategy. A NP is not affecting a strategic goal if the achievement of the goal, no matter *how* it is achieved, does not allow for a violation of the NP. If, according to how a goal is achieved, the NP can be violated or respected, the goal needs to be elaborated for compliance.

3. *Choose realizing goals.* A compliance goal appears in an actor's rationale because the actor must be compliant with some NP. A top-level compliance goal realizes a NP — i.e., there is a realization relation from that goal to the NP, meaning that satisfying that goal satisfies the NP, and indicates compliance with that NP. The realization relation results in practice in that the NP that the goal realizes is a constraint on the definition of the solution for that goal: any solution for the goal must be such that the realization relation is not violated, i.e., that it is not dropped by a solution because that solution satisfies the goal in a way that violates the NP.

4. *Refine compliance goals.* A compliance goal must be refined in such a way that every alternative refinement thereof must not violate the NP that this compliance goal realizes. Subgoals in the refinements must be chosen in such a way that their joint satisfaction does not violate the NP.

5. *Operationalize compliance goals.* If $g$ is a compliance goal for a given NP, a solution for it has to be sought by refining/decomposing $g_k$ onto subgoals, then finding tasks the execution of which satisfies the subgoals. If $g$ is AND-refined/decomposed onto subgoals $g_1$ and $g_2$, then $g$ is compliant if and only if the conjunction of the two subgoals is compliant: neither the satisfaction of $g_1$ alone, nor of $g_2$ alone is good enough to ensure that $g$ is satisfied. In general terms, the solution $sol(g)$ for the compliance goal $g$ is the conjunction of its leaf tasks.

*Outcomes.* Definition III.17 stated the compliance problem in argumentation terms as: $Spec(R), C \mathrel{\mid\!\sim}_C L$. After the application of the described process, we have that: $L$ is the input model of the law we want to address; $R$ is the initial requirements set, refined in the steps 3 and 4; $Spec(R)$ amounts in the solution for $R$ - i.e., the sets of tasks that operationalize the compliance goals; $C$ contains the set of realization relations. The compliance condition expressed by the $\mathrel{\mid\!\sim}_C$ relation can't however be formally proved. Rather, it has to be argued by providing evidence that the compliance modeling choices are reasonably acceptable.

**Phase 2. Argumentation**  The compliance of the framework is deduced form the dialectical tree $T(a(\phi))$. The tree is primarily formed by the arguments collected during phase 1 of the process. For example, if a goal delegation resulted from a consideration on the abilities of the delegatee in fulfilling the goal, such considerations are added to the argumentation as a support that that modeling choice. Phase 2 consists in debating about the modeling choices. Three types of argument can be given:

(i) Arguments in favor of the modeling choices; i.e., arguments that confirm the conclusions of the argumentation tree built during the development of the solution (phase 1).

(ii) Arguments in favor of for changing or replacing some parts of the solution; in this case, such arguments are added to the argumentation tree, and a modeling activity is undertaken again to change $R$ into an $R'$.
(iii) Arguments that reject partly or completely the conclusions of the argumentation tree; in this case, a new solution for $L$ has to be searched.

**Phase 3. Verification** The dialectical tree can be traversed for making conclusions form it by means of a justification process as described in Definition III.16. The process starts from the leaves and makes conclusions on the acceptability of the roots, which are basically the compliance hypotheses (as of step 3 of phase 1). If the justification process allows for deducing acceptance of the compliance hypotheses, then compliance is demonstrated on an argumentation basis. If compliance can be demonstrated for every NP, we say that the requirements $R$ are compliant with the law $L$. If new arguments come in, the validity of a conclusion is not ensured anymore, and the process has to be iterated.

## 6 Example

Figure 2 depicted an example of a compliance solution for a certain NP. Why that model is actually a solution? Is it really enough to claim compliance with the NP? Should the solution be revised?

HIPAA section §45CFR164.314 states the general organizational requirements for privacy and security that every Covered Entity (CE) has to respect. *Who* is a CE is defined in paragraph §45CFR164.501. With regard to this definition, the Hospital: $a_1$ — "deals with patients' PHI"; $a_2$ — "furnishes a business health care service"; $a_3$ — "is a private company". Without further going into the details, from the argumentation $T_A$, comprised by the arguments $a_1, a_2, a_3$, one can argue the conclusion that the Hospital is a CE. So, the NP from paragraph §502a: $np_h$ = *A CE must keep patients' PHI closed* is applicable to the Hospital, and it has to comply with it.

With respect to $np_h$ the goal $g_1$ (reported in Figure 3) can be argued to be a possible violation point: $a_4$ — *if an operator retrieves EPR data, then he is a possible leak point in the system*; this is an argument *in favor of* the applicability of $np_h$ to $g_1$. A possible attack to $a_4$ is $a_5$ — *data into the EPR is anonymous, so there is no need to keep it closed*; this argument rebuts $a_4$. Differently, $a_5$ is undercut by $a_6$ — *at the registration, EPR data must be associated with the patient's identity*. The arguments and attacks between them form a dialectical tree $T_B$, as explained in Definition III.16, which also indicates how to determine the accepted or rejected status of each argument in the dialectical tree.

On the way for finding a solution for $g_1$, we could argue that: $a_7$ — *If PHI is only processed electronically, no disclosure can happen through other channels*; and $a_8$ — *Preventing unauthorized access to PHI prevents its unauthorized disclosure*; then we argue a goal like $g_2$ (*Provide regulated EPR access*) could allow for full compliance of $gp_1$ with $np_h$. So, from the dialectical tree $T_C = \langle a_7 \xrightarrow{A} a_8 \rangle$ we hypothesize that $g_1$ can be a realizing goal for $np_h$. We add the "realization" relation between $g_1$ and $np_h$ to the set of compliance assumptions $C$; so, $C = \{sol(g_1) \to np_h\}$. This hypothesis will guide the consequent modeling of the compliance solution $sol(g_1)$. If successful, it will end up in the operationalization of $g_1$ in terms of tasks.

If we want to make compliant the whole process that underlies the *Book service* goal, we have to refine $g_1$. In the example of Figure 3, we have made some further addition to both the model and the dialectical tree. We added $a_9$ = *The booking service can be compliant, if we constrain the use of patient PHI* to $T_C$, so that the realizing goal is desired to be the higher-level goal *Book service*; also, we add the same argument to $T_D$, which motivates the decomposition of $g_1$ into $g_2$ and $g_3$. Similarly, the modeling choice to delegate $g_3$ to the BS actor is added as an argument to $T_C$, and so on. Arguments like these form the dialectical tree $T_D$, which is the justification for goal decompositions and delegation of the model in Figure 3. The goals added to the model due to such arguments form new requirements in $R$.

The modeling process ends when a full operationalization of the compliance goals is found (step 5) — i.e., when a $Spec(R)$ is found. For the goal *Book service through EPR* in Figure 3, the BS has to ask the personal information to the patient (task $t_7$), access the EPR system (task $t_5$) and insert the patient's data (task $t_5$). Some of these tasks do not raise any compliance issue — for example, $a_{10}$ = *asking the data to the patient, per se, does not result in data disclosure to others*. In this case, the argument is added to $T_D$. Other tasks, such as *Access EPR system*, can be attacked w.r.t. their capability to remain compliant. For example, it is easy to argue that, $a_{11}$ = *while accessing the EPR system, the BS can perform unauthorized transmissions*. However, due to the argumentation (in $T_C$) for the And-decomposition of the goal $g_1$ (*Book service*), which undercuts $a_{11}$, we have that EPR can only be accessed in ways regulated by the Hospital. For readability concerns, we have highlighted this situation in the picture by means of the goal dependency between the task and the *Constrain User data access* of the Hospital actor. The meaning of the dependency is that the task can be argued to be compliant due to the same argumentation that makes compliant the operationalization of the Hospital goal. It can be proved that for each of the leaf tasks there is an argumentation saying that either (a) the task does not raise compliance issues, or (b) the task is compliant with the normative proposition due to an argumentation. Going back to Figure 1, the solution for Hospital's goal $g_1$ ($sol(g_1)$) consists in an and-decomposition of $g_1$ into $g_2$ and $g_3$, in the delegation of $g_3$ to the BS, and in the and-decomposition of $g_2$ into $t_1, t_2, t_3, t_4$ and of $g_4$ into $t_5, t_6, t_7$. The set $t_1, ..., t_7$ form $Spec(R)$, while $(t_1, ..., t_7) \rightarrow np_h$ form $C$.

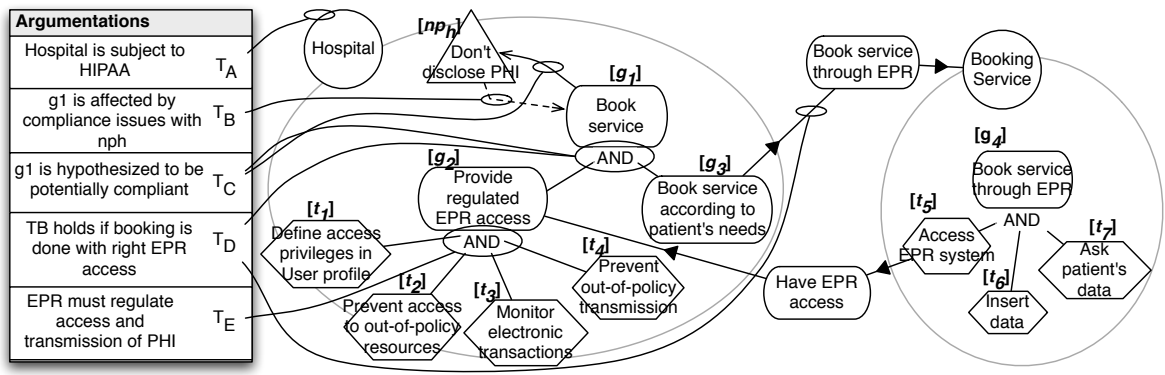

**Figure 3:** An example of how the dialectical trees support the compliance evaluation provided by a *Nómos* model.

The model in Figure 3 proposes a solution for the problem of the compliance of $g_1$ with $np_h$. As said, the proposed solution claims for an argument-theoretic acceptability rather than a truth-theoretic validity: further arguments can be raised, challenging for example the correctness or completeness of the model, and requiring to iterate again the process until an agreement on the model's acceptability is found. On the other hand, if, over the iterations, $R$ changes to an $R'$, the non-monotonicity of $\vdash_C$ causes any acceptable solution on $R$ to be not necessarily acceptable on $R'$. For example, if we add to $g_2$ another task – let say, $t_8 =$ *Save monitoring log*, then the addition of this new requirement can compromise the solution. In fact, saving the monitoring log can be argued to create further privacy issues, since the log can be in turn be disclosed. In the end, we have shown that a compliance solution can be expressed by the elements in the picture: a model of $L$; some requirements $R$ and their specification $Spec(R)$; some compliance assumptions $C$; and the dialectical tree associated to the model, which represent justification of acceptability for the compliance solution.

## 7 Related Works

Allen [1] firstly pointed out the possibility to formalize law as a logical theory, in order to be able to make conclusions about the consequences of the law. Worth mentioning in particular Stamper's LEGOL [15], a LEGally Oriented Language aimed at expressing a legislation by means of formal rules, suitable to be processed by computers. The idea behind LEGOL was that legislations shape the behaviour of information systems: regardless on *how* information systems are implemented, they perform within the boundaries created by laws.

In recent years, law compliance has increasingly been addressed in relation with requirements engineering. Anton and Breaux have developed a systematic process, called semantic parameterisation, which consists of identifying in legal text restricted natural language statements (RNLSs) and then expressing them as semantic models of rights and obligations [5] (along with auxiliary concepts such as actors and constraints).

Darimont and Lemoine have used KAOS as a modelling language for representing objectives extracted from regulation texts [6]. Such an approach is based on the analogy between regulation documents and requirements documents. Ghanavati et al. [9] use GRL to model goals and actions prescribed by laws. This work is founded on the premise that the same modelling framework can be used for both regulations and requirements. Likewise, Rifaut and Dubois use $i*$ to produce a goal model of the Basel II regulation [12].

## 8 Conclusions

We have proposed a formal framework for regulatory compliance through which one can establish that a set of requirements complies with a given law. The framework is non-monotonic in that new compliance assumptions may invalidate compliance, and relies on argument-theoretic – rather than truth-theoretic – concepts. The framework is instantiated by adopting a particular modeling language for laws and requirements (*Nómos*) and a systematic process is presented for taking a given set of requirements $R$ and a law $L$ and refining/extending $R$ to generate a new set of requirements $R'$ that are compliant with $L$, assuming a set of compliance assumptions $C$.

Ongoing work involves the application of the framework to several case studies in order to gain additional insight into the relevance of the testable hypotheses. We are working on the refinement of the framework as described in Section IV and studying whether argumentation with preferences [11] can be relevantly used in Nomos. We intend to provide a complete software environment for the modeling of norms and requirements, along with the component that will handle argumentation.